\documentclass[prl,twocolumn,superscriptaddress,aps,longbibliography]{revtex4-1}
\usepackage{natbib}
\usepackage{graphicx}
\usepackage{enumitem}
\DeclareGraphicsExtensions{.pdf,.png,.jpg}
\usepackage{array}
\usepackage{amsmath} 
\usepackage{amsthm} 
\usepackage{amssymb}    
\usepackage{latexsym}
\usepackage{amsfonts}
\usepackage{mathrsfs}
\usepackage{color}
\usepackage{bbold}

\usepackage{subfigure}
\usepackage[colorlinks=true, urlcolor=blue,citecolor=blue,linkcolor=blue]{hyperref}

\begin{document}
\newcommand{\bra}[1]{\left< #1\right|}   
\newcommand{\ket}[1]{\left|#1\right>}
\newcommand{\abs}[1]{\left|#1\right|}
\newcommand{\ave}[1]{\left<#1\right>}
\newcommand{\Tr}{\mbox{Tr}}
\renewcommand{\d}[1]{\ensuremath{\operatorname{d}\!{#1}}}
\renewcommand\qedsymbol{$\blacksquare$}

\title{Necessary and sufficient condition for joint measurability}
\author{Jeongwoo Jae}
\affiliation{Department of Physics, Hanyang University, Seoul, 04763, Republic of Korea}
\author{Kyunghyun Baek}
\affiliation{School of Computational Sciences, Korea Institute for Advanced Study, Seoul, 02455, Republic of Korea}
\affiliation{Asia Pacific Center for Theoretical Physics, Pohang, 37673, Republic of Korea}
\author{Junghee Ryu}
\affiliation{Centre for Quantum Technologies, National University of Singapore, 3 Science Drive 2, Singapore 117543}
\author{Jinhyoung Lee}
\email{hyoung@hanyang.ac.kr}
\affiliation{Department of Physics, Hanyang University, Seoul, 04763, Republic of Korea}
\begin{abstract}
In order to analyze joint measurability of given measurements, we introduce a Hermitian operator-valued measure, called $W$-measure, such that it has marginals of positive operator-valued measures (POVMs). We prove that ${W}$-measure is a POVM {\em if and only if} its marginal POVMs are jointly measurable. The proof suggests to employ the negatives of ${W}$-measure as an indicator for non-joint measurability. By applying triangle inequalities to the negativity, we derive joint measurability criteria for dichotomic and trichotomic variables. Also, we propose an operational test for the joint measurability in sequential measurement scenario.
\end{abstract}
\maketitle

{\em Introduction.---}Quantum physics does not allow to exactly and simultaneously specify two observable properties when their measurements are incompatible, as implied by complementarity principle~\cite{Bohr28}. The incompatibility can be discussed in terms of the non-commutativity between the observable operators: If their operators commute, measurements are compatible so that they can be jointly performed~\cite{Heisenberg1927}. This holds for the measurements represented by observable operators, or equivalently, (von-Neumann) projection-valued measures (PVMs). The discussions on joint measurability (JM) need to be generalized to positive operator-valued measures (POVMs) from PVMs~\cite{Busch86}.

The complementarity principle implies the trade-off relations between the certainties of observed quantities. {JM} is employed to characterize the trade-off relations; the wave-particle duality~\cite{Englert96,Liu09} and the error-disturbance~\cite{Busch2013,*Busch2014}. The non-joint measurability (non-JM) also attracts attentions since it is intimately related to the violation of Bell inequality~\cite{Andersson05,Son05,Wolf09}, the quantum steering~\cite{Wiseman07}, the temporal steering~\cite{Karthik15,Uola2018}, and the quantum contextuality~\cite{OH2013}. In particular, the steerability is identified by a test whether the local measurements are jointly measurable for a given entangled state~\cite{Uola2014,*Quintino2014}.

The JM was studied for a pair of discrete~\cite{Busch86,Carmeli2012} or continuous variables~\cite{Teiko2014,Kiukas17}. It was also studied for a triad of dichotomic variables~\cite{OH2013}. These studies were focused on POVMs of mutually unbiased bases, while general bases were considered for a pair of dichotomic variables ~\cite{Busch86,Sixia10}. There were proposals to realize joint measurements by a sequential~\cite{Carmeli2011} and an adaptive~\cite{Uola2016} methods. The numerical analyses were made by semidefinite programming~\cite{Wolf09,Bavaresco17}. As increasing the number(s) of measurements and/or outcomes, however, the identification of JM becomes a hard problem, since the geometric complexity grows significantly in the operator space~\cite{Byrd03}. In fact, the related geometric problems called generalized Fermat-Toricelli point~\cite{OH2013} or $1$-median problem is difficult to find the exact solution(s) by any known algorithms~\cite{Megiddo84}.

Operational tests of JM are nontrivial problems. Non-JM is unable to test by its realization in quantum physics. For an operational test, we employ the indirect method based on operational quasiprobability~\cite{Ryu2013,Jae2017}, which is determined by the statistics from measurements, similarly to the device-independent approaches~\cite{Clauser81,Gallego10}. Negative operational quasiprobability is an indicator of nonclassicality such as entanglement~\cite{Ryu2013}, the violation of macrorealism~\cite{Jae2017}, and the measurement-selection context~\cite{Ryu2018}.

In this Letter, we propose an analyzing tool to examine both JM and non-JM, applicable for general system. To this end, we introduce a Hermitian operator-valued measure, called ${W}$-measure, such that its marginals are POVMs to test their JM. We provide a theorem that ${W}$-measure is a POVM {\em if and only if} its marginal POVMs are jointly measurable. It turns out that the negativity of $W$-measure can be used as an indicator to non-JM. The negativity is represented by distance measure, satisfying triangle inequality, and this property is core to obtain joint measurability criteria. We first apply this approach to the measurements of dichotomic variables to reproduce the known JM criteria. We also derive a new JM criterion for trichotomic variables on a qubit. Finally, we discuss an operational test for non-JM involving a sequential measurement.

{\em Joint measurability and ${W}$-measure.---}Suppose there is a measurement, represented by a POVM $J=\{\hat{J}_{ij}\}$; $\hat{J}_{ij}$ are positive (semidefinite) operators, $\hat{J}_{ij} \geq 0$, and satisfy the completeness relation, $\sum_{ij} \hat{J}_{ij} = {{\mathbb{1}}}$. The pair of indices $(i,j)$ denote the measurement outcomes. Let $A$ and $B$ be marginals of $J$ such that $A=\lbrace\hat{A}_i = \sum_j \hat{J}_{ij}\rbrace$ and $B=\lbrace \hat{B}_j = \sum_i \hat{J}_{ij}\rbrace$. The marginals $A$ and $B$ are said jointly measurable and $J$ is referred to as a joint measurement when the marginals represent the two measurements, i.e., they are POVMs; $A$ satisfies $\hat{A}_i \geq 0$ and $\sum_i \hat{A}_i = {{\mathbb{1}}}$. Similarly $B$ does. The backward problem, finding a joint POVM $J$ for given POVMs $A$ and $B$, is generally hard to solve. We exploit a $W$-measure to attack the backward problem.

We define a ${W}$-measure by the set of {\em Hermitian} operators in $D$-dimensional Hilbert space, $W=\{\hat{W}_{ij}\}$, such that its elements satisfy the completeness $\sum_{i,j} \hat{W}_{ij} = {{\mathbb{1}}}$ and are given by
\begin{eqnarray}\label{eq:Wmeasure}
\hat{{W}}_{ij}= \hat{C}_{ij} + \frac{1}{d}\left( \hat{A}_i-\sum_{j=1}^{d}\hat{C}_{ij}\right) + \frac{1}{d}\left(\hat{B}_j-\sum_{i=1}^{d}\hat{C}_{ij}\right),
\end{eqnarray}
where $i$ and $j$ denote $d$ outcomes. The $W$-measure is composed of the three POVMs. POVMs $A=\{\hat{A}_i\}$ and $B=\{\hat{B}_j\}$ represent two measurements to test JM. POVM $C=\{\hat{C}_{ij}\}$ does a conjunction measurement, which we adjust in the test of JM. As defined in an operational way, the $W$-measure can be constructed through the tomography~\cite{Lundeen09}. Note that the test POVMs $A$ and $B$ are the marginals of the $W$-measure, $\sum_j \hat{W}_{ij} = \hat{A}_i$ and $\sum_i \hat{W}_{ij} = \hat{B}_j$. This implies that the $W$-measure becomes a joint POVM of $A$ and $B$ if its elements are all positive, $\hat{W}_{ij} \geq 0, \forall i,j$. One of our main results is thus stated by the following theorem.

{\bf Theorem 1.} POVMs $A$ and $B$ are jointly measurable, {\em if and only if} the $W$-measure is a POVM for some conjunction POVM $C$.

{\bf Corollary 1.} There exists some conjunction POVM $C$ if the $W$-measure is a POVM.

The proofs for the theorem and the corollary are given in the Supplementary material~\cite{Note1}. Theorem 1 shows that the positivity of ${W}$-measure plays a role of indicator that the given POVMs $A$ and $B$ are jointly measurable. Corollary 1 enables to release the positivity condition of conjunction $C$, which we use in theorem 2.

{\em Negativity of ${W}$-measure.---}As the theorem 1 states, the existence of the negative eigenvalues of $W$-measure indicates non-JM. We quantify the {\em negativity} as
\begin{equation}\label{eq:negativity}
{\cal N} := \frac{1}{D}\sum_{i,j} \left\Vert | \hat{W}_{ij} | - \hat{W}_{ij} \right\Vert \ge 0,
\end{equation}
where $\Vert \hat{X}- \hat{Y} \Vert = \text{Tr} | \hat{X} - \hat{Y} |$ is the trace norm between operators $\hat{X}$ and $\hat{Y}$, and $| \hat{X} | = \sqrt{\hat{X}^\dag \hat{X}}$. For Hermitian operator $\hat{W}$, the trace norm $\left\Vert |\hat{W}| -  \hat{W} \right\Vert = \sum_{k=1}^D (|\lambda_k| - \lambda_k)$, where $\lambda_k$ are the eigenvalues of $\hat{W}$. In particular, $\left\Vert |\hat{W}| -  \hat{W} \right\Vert = 0$ for a positive operator $\hat{W} \geq 0$. Thus, the minimum ${\cal N} = 0$ is attained if and only if the $W$-measure is a POVM.

We deal with a $W$-measure by introducing a set of differential operators $\Theta = \{\hat{\Theta}_{ij}\}$, which are Hermitian and satisfy the constraints $\sum_{i}\hat{\Theta}_{ij}=\sum_{j}\hat{\Theta}_{ij}={{\mathbb{1}}}/d$. With ${\Theta}$, the $W$-measure is represented by
\begin{equation}
{\hat{W}_{ij}}= \frac{1}{d}\left( \hat{A}_i+\hat{B}_j \right) - \hat{\Theta}_{ij},
\end{equation}
where the marginality is preserved now by the $\Theta$ constraints. This representation arises due to corollary 1. In other words, we always find a conjunction POVM $C$ for a positive $W$-measure and we can concentrate on the differential operators $\Theta$. Theorem 1 is rephrased in terms of the negativity, ${\cal N} = {\cal N}(\Theta)$:

{\bf Theorem 2}. POVMs $A$ and $B$ are jointly measurable if and only if the negativity ${\cal N}(\Theta) = 0$ for some $\Theta$.

The proof for theorem 2 is given in the Supplementary material~\cite{Note1}. Theorem 2 translates the problem of JM, usually attacked by using a semidefinite program, to the minimization of a single real-valued function ${\cal N}(\Theta)$, over $(d-1)^2$ independent differential operators in $\Theta$ due to the $\Theta$ constraints. To parameterize $\Theta$, we take a generalized Bloch representation~\cite{JLee03},
\begin{eqnarray}\label{eq:newform}
\hat{\Theta}_{ij} = \frac{1}{d^2}\left(\theta_0^{ij} {{\mathbb{1}}} + \sum_{k=1}^{D^2-1} \theta^{ij}_k \hat{\gamma}_k \right),
\end{eqnarray}
where generalized Bloch real vectors $\vec{\theta}_{ij} = (\theta^{ij}_1, \theta^{ij}_2, \cdots, \theta^{ij}_{D^2-1})$, and the basis set $\{\hat{\gamma}_k\}_{k=0}^{D^2-1}$ is composed of the identity ${{\mathbb{1}}}=\hat{\gamma}_0$ and traceless Hermitian operators $\hat{\gamma}_k$, satisfying the orthogonality $\text{Tr} \hat{\gamma}_k \hat{\gamma}_l = D \delta_{kl}$, on the Hilbert-Schmidt space.

We first show that our method reproduces JM criteria for dichotomic variables~\cite{Busch86,Sixia10}. The bases of $2$-outcome measurements are given in the Bloch representation by $\hat{A}_i = ({{\mathbb{1}}} + \omega^i \vec{a} \cdot \vec{\sigma})/2$ for $i=1,2$, where $\vec{a}$ is a Bloch vector representative of POVM $A$ with $|\vec{a} | \le 1$, $\omega=-1$, and $ \vec{\sigma}= (\hat{\sigma}_x, \hat{\sigma}_y, \hat{\sigma}_z)$ with Pauli operators $\hat{\sigma}_i$. For the POVMs, $A=\{\hat{A}_i\}$ and $B=\{\hat{B}_j\}$, the minimized negativity over $\Theta$ is given by
\begin{equation}\label{eq:functionN}
{\cal N}_\text{min}= \max\left( \frac{1}{4}\sum_{i.j=1}^2|\vec{a}+\omega^{i+j}\vec{b}| - 1, 0\right),
\end{equation}
where $\vec{a}$ and $\vec{b}$ are the Bloch vector representatives of $A$ and $B$, respectively. This is proved in Supplementary material~\cite{Note1} by using triangle inequalities between trace distances.

Eq.~\eqref{eq:functionN} implies that the condition, ${\cal N}_\text{min} = 0$, is reduced to the well-known criterion~\cite{Busch86}: $|\vec{a}+\vec{b}| + |\vec{a}-\vec{b}| \le 2$. Also, the minimized negativity is proportional to the ``incompatibility'' of generalized uncertainty relation~\cite{Busch2014}. These results hold for outcome-unbiased POVMs, by which we mean that they have no preference to a particular outcome, when averaged over all pure states: $\bar{A}_i = \frac{1}{\Omega} \int d \psi \langle \psi | \hat{A}_i | \psi \rangle = \frac{1}{2}, \forall i$.

\begin{figure}[t]
	\centering
	\includegraphics[width=0.49\textwidth]{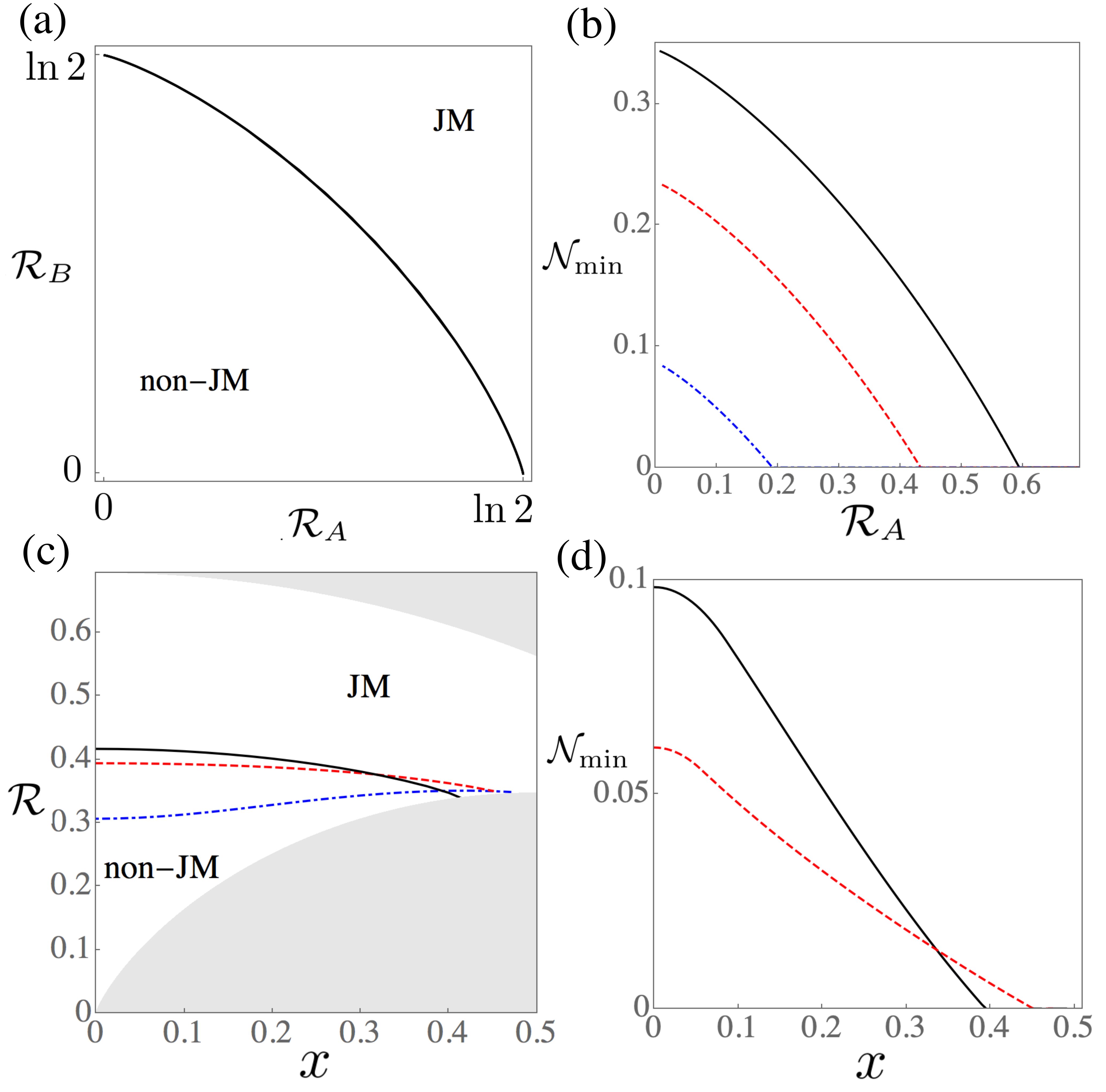}
	\caption{(a) Boundary between JM and non-JM regions on the space of the unsharpnesses ${\cal R}_{A,B}$ for test POVMs $A$ and $B$, which are $2$-outcome unbiased and satisfy $\vec{a} \cdot \vec{b} =0$. The test POVMs enter the JM region as the unsharpness increases. (b) Minimized negativity ${\cal N}_\text{min}$ as a function of ${\cal R}_A$ for ${\cal R}_B= 0.2$ (solid), $0.4$ (dashed), and $0.6$ (dot-dashed) in case of (a). As the unsharpness ${\cal R}_A$ increases, the negativity ${\cal N}_\text{min}$ decreases down to zero, where the test POVMs are jointly measurable. The smaller ${\cal R}_B$ is, the larger ${\cal R}_A$ is required to enter the JM region. (c) Boundaries between JM and non-JM on the space of the unsharpness ${\cal R}$ and the outcome-biased degree $x$ for $\phi=\pi/2$ (solid), $\pi/3$ (dashed), and $\pi/6$ (dot-dashed). The test POVMs $A$ and $B$ have the same degrees of outcome bias $|a_0|=|b_0|=x$ and the same unsharpnesses ${\cal R}_{A,B} = {\cal R}$, and their Bloch vectors are relatively oriented with the angle $\phi$. No POVMs $A$ and $B$ exist in the shaded regions. For large $\phi=\pi/2$, ${\cal R}$ decreases as $x$, whereas it increases as $x$ for small $\phi=\pi/6$. The boundaries intersect in large $x \gtrsim 0.4$. (d) Minimized negativity ${\cal N}_\text{min}$ as a function of $x$ for $\phi=\pi/2$ (solid) and $\pi/3$ (dashed) in case of (c). The negativities decrease as $x$. On the other hand, they intersect around $x=0.35$, and the test POVMs with $\phi=\pi/2$ enter the JM region earlier in $x$ than $\phi=\pi/3$.}
\label{fig:OHrp}
\end{figure}

Outcome-biased POVMs are given by $\hat{A}_i = [(1\pm |a_0|){\mathbb{1}} + \omega^i \vec{a} \cdot \vec{\sigma}]/2$, where $|a_0| + |\vec{a}| \le 1$ and $|a_0|$ is the degree of bias in outcomes as $\bar{A}_i = (1+\omega^i a_0)/2$. Two biased POVMs $A$ and $B$ are jointly measurable~\cite{Sixia10} when
\begin{eqnarray}\label{eq:OH}
\left( 1- F_A^2 - F_B^2  \right) \left( 1 - \frac{a_0^2}{F_A^2} - \frac{b_0^2}{F_B^2} \right) \le \left( \vec{a} \cdot \vec{b} - a_0 b_0 \right)^2,
\end{eqnarray}
where $F_A = \frac{1}{2} \sum_{f=\pm 1} \sqrt{(1+ f a_0)^2 - |\vec{a}|^2}$ and $F_B$ is similarly given. In terms of the negativity, we numerically examine the boundaries of JM/non-JM as in Fig.~\ref{fig:OHrp}.

The unsharpnesses of POVMs are crucial factors, which determine JM. As blunted by noises, the measurements become unsharp and change their JM. For a measure of unsharpness~\cite{Baek16,*Baek162}, we employ the unsharpness entropy, defined by
\begin{eqnarray}
{\cal R}_A := \frac{1}{D} \text{Tr} \hat{R}_A,
\end{eqnarray}
where the entropy operator $\hat{R}_A = -  \sum_{i=1}^{d} \hat{A}_i \ln \hat{A}_i$. The unsharpness entropy is equal to the average of the entropy operator over all pure states: ${\cal R}_A = \frac{1}{\Omega} \int d \psi \langle \psi | \hat{R}_A | \psi \rangle$. POVM $A$ is a PVM if ${\cal R}_A = 0$, the least unsharp, while it is a purely random measurement if ${\cal R}_A = \ln d$, the most unsharp. In general, $0 \le {\cal R}_A \le \ln d$. The unsharpness entropy is monotonically related with other measures of unsharpness (or sharpness) formulated with the linear entropy~\cite{Busch86,Busch96} and the distinguishability~\cite{Sixia10,Englert96}. Fig.~\ref{fig:OHrp} presents the role of the unsharpnesses on the negativity $\cal N$ and the joint measurability.

{\em Joint measurability of trichotomous variables.---}In the previous section we saw that the JM of given measurements depends on their unsharpness, the relative orientation of their Bloch vectors, and their degrees of outcome bias. We are interested in this kind of properties for trichotomous variables, how large unsharpness are required to be jointly measurable. For trichotomous measurements on a qubit, this is not obvious; No PVMs are of $3$ outcomes, and trichotomous POVMs are basically unsharp on a qubit.

\begin{figure}[t]
	\centering
	\includegraphics[width=0.22\textwidth]{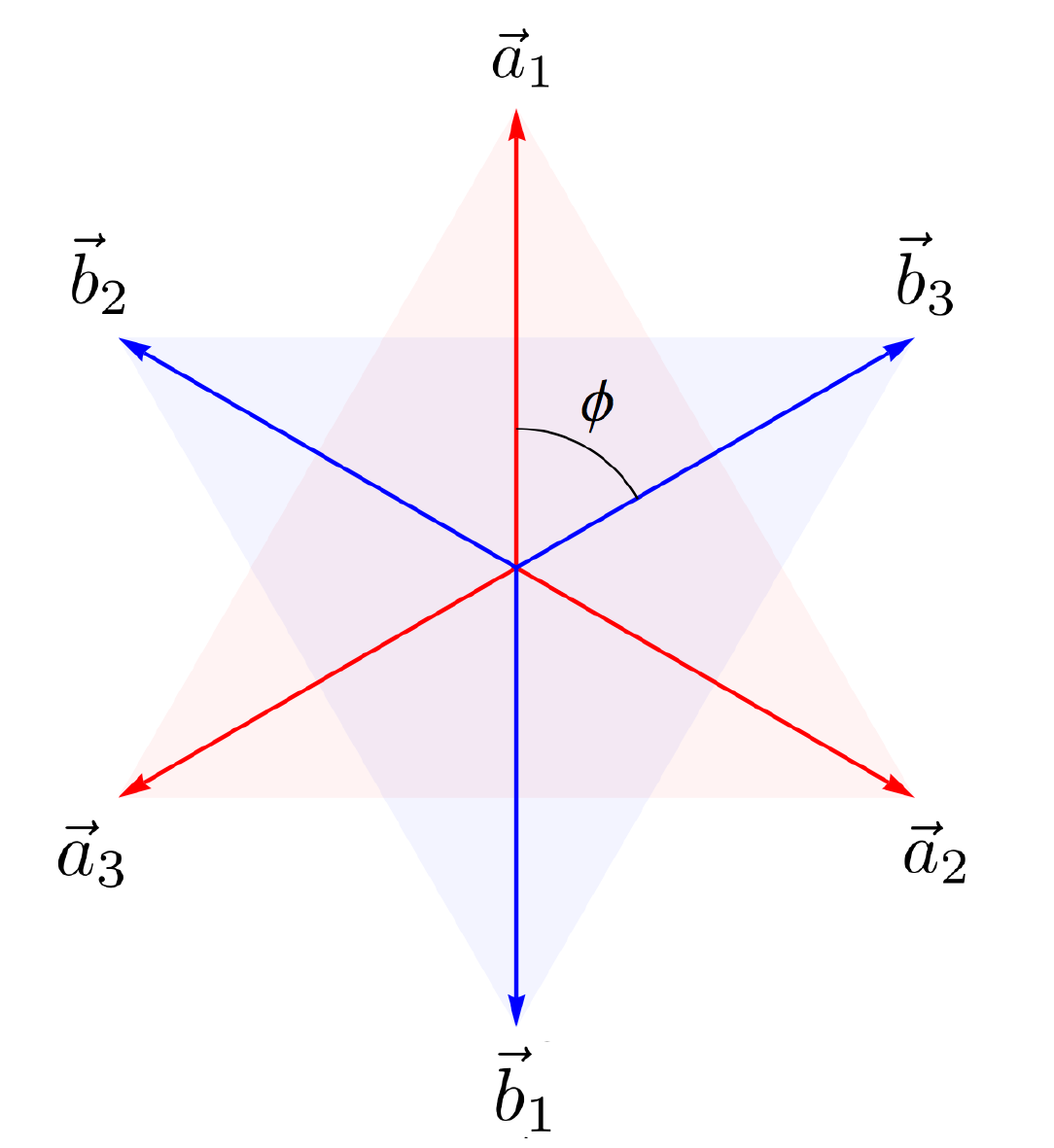}
	\caption{Geometry of the unit Bloch vectors ${\vec{a}}_i$ and ${\vec{b}}_j$ for the trichotomous POVMs $A$ and $B$, respectively. Each set of Bloch vectors forms a triangle. The triangles are assumed in the same plane, and the $\phi$ denotes the relative angle between them. For integer $n$, $A$ and $B$ are commutative if $\phi=2\pi n/3$, whereas they are the most non-commutative if $\phi=(2n+1)\pi/3$.}
	\label{fig:geometry}
\end{figure}
$3$-outcome unbiased measurements on a qubit are represented by POVMs, $\hat{A}_i = ({\mathbb{1}}+\vec{a}_i \cdot \vec{\sigma})/3$ with $d=3$, where $|\vec{a}_i| \le 1$. The completeness of $A$ demands that $\sum_i \vec{a}_i = \vec{0}$ and the $3$ vectors $\vec{a}_i$ form a triangle in the Bloch space (contrary to a $2$-outcome POVM, whose Bloch vectors form a line segment). Then, we associate a triangle with a POVM in regards of shape, size, and orientation. We assume that the triangles of POVMs are equilateral and in the same plane (see Fig.~\ref{fig:geometry}). The two triangles compose a hexagram when $\phi = \pi/3$. The POVMs are commutative and mutually equivalent when $\phi=2\pi/3$, while the outcomes are shifted. JM of $A$ and $B$ then depends on their relative orientation as well as their unsharpnesses.
\begin{table}[t]
\caption{\label{table:3o2d}Threshold unsharpness entropies ${\cal R}_{\text{th}}$ and sharpnesses $\mu_{\text{th}}$ of the two $3$-outcome measurements such that the test POVMs are jointly measurable when ${\cal R}\geq {\cal R}_{\text{th}}$ and $\mu\leq\mu_{\text{th}}$.}
\begin{ruledtabular}
\begin{tabular}{ccccc}
~~~~~$\phi$ &$0$ &$~~~\pi/6$ & $~~~{\pi}/{4}$ & $~~~{\pi}/{3}$\\
 \colrule
 ~~~~~${\cal R}_{\text{th}}$& $\ln (3/2)$  & $0.608989$ & $0.641283$ & $0.651238$\\
~~~~~$\mu_{\text{th}}$ & $1$ & $0.896575$ & $0.873498$ & $\sqrt{3}/2$
\end{tabular}
\end{ruledtabular}
\end{table}

\begin{figure}[t]
	\centering
	\includegraphics[width=0.47\textwidth]{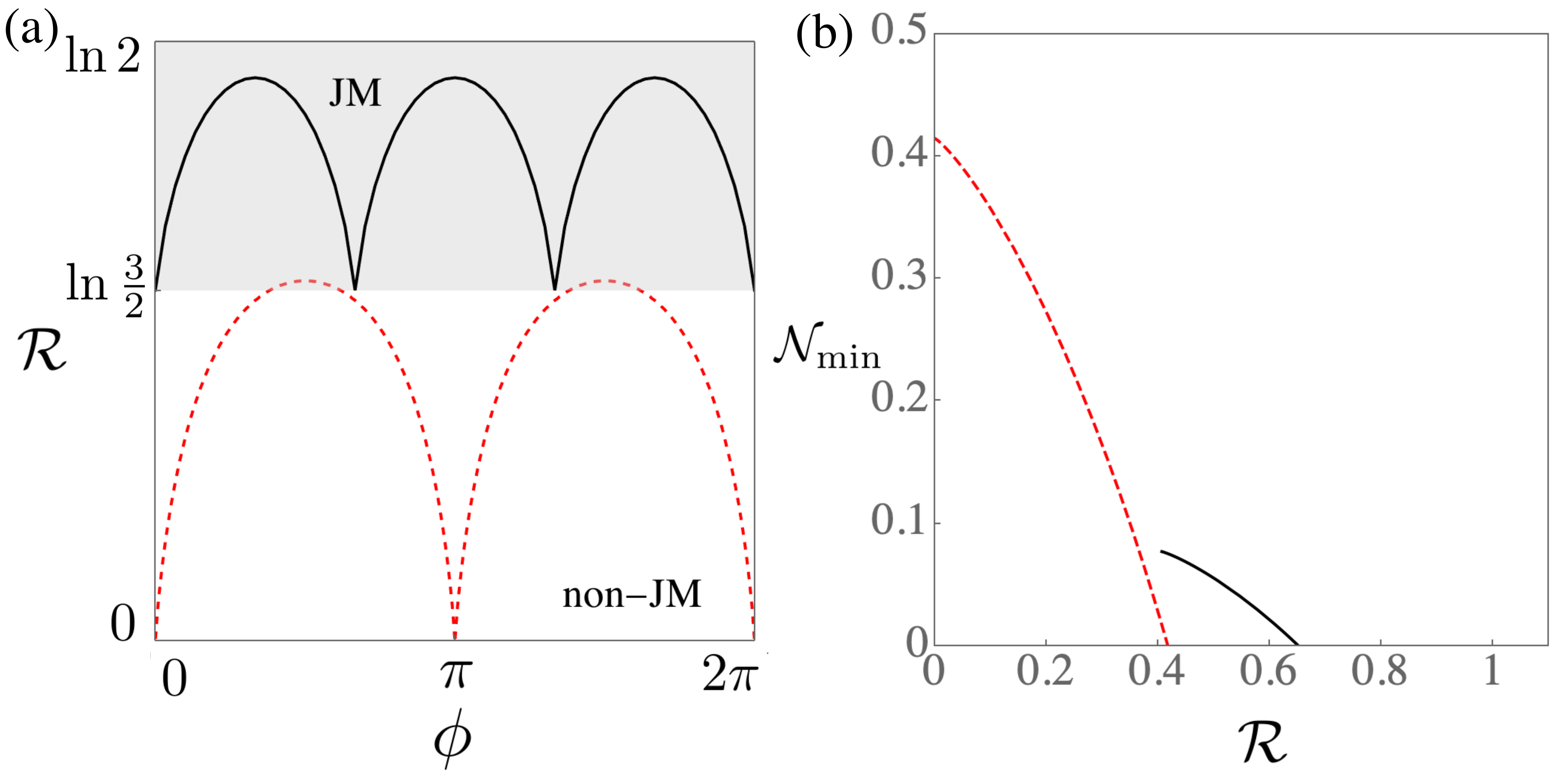}
	\caption{(a) Joint measurability (JM) criteria for the trichotomous POVMs $A$ and $B$ on the space of relative angle $\phi$ and unsharpness ${\cal R}$. The solid line is the threshold unsharpness, ${\cal R}_\text{th}$. This is a boundary, which separates the JM and the non-JM regions. The region is shaded where the trichotomous POVMs exist. The boundary alternates as $\phi$. The threshold unsharpeness ${\cal R}_\text{th}=\ln (3/2) \approx 0.41$ is the minimum at $\phi=2\pi n/3$ for an integer $n$, where $A$ and $B$ are commutative. ${\cal R}_\text{th} \approx 0.65$ is the maximum at $\phi=(2n +1) \pi /3$, where $A$ and $B$ are the most non-commutative as in Fig.~\ref{fig:geometry}. These results of the trichotomous variables are compared to those of the dichotomic variables. For the dichotomic variables, the minimum ${\cal R}_\text{th} = 0$ when $A$ and $B$ are commutative, and the maximum ${\cal R}_\text{th} \approx 0.42 $ for the most non-commutative $A$ and $B$ with $\vec{a}\cdot\vec{b}=0$~\cite{Busch86,Sixia10}. The boundary of the JM and non-JM regions are represented by the dashed line. (b) Minimized negativity ${\cal N}_\text{min}$ as a function of the unsharpeness ${\cal R}$ with the given $\phi = \pi/3$ for the trichotomous variables (solid) and the dichotomic variables (dashed).}
	\label{fig:RPthree}
\end{figure}

As in $2$-outcome POVMs, we minimize the negativity ${\cal N}(\Theta)$ over the set of free differential operators $\Theta$ for a given angle $\phi$, obtaining
\begin{equation}\label{eq:threN}
{\cal N}_\text{min} = \max \left( \frac{1}{9}\sum_{i,j=1}^3 |\vec{a}_i+\vec{b}_j - \vec{\theta}^s_{ij}| - 1, 0 \right),
\end{equation}
where $\vec{\theta}^s_{ij} = \vec{a}_{2(i+j)} + \vec{b}_{2(i+j)}$ are the Bloch vectors of $\hat{\Theta}_{ij}$ that minimizes the negativity (and the subscripts are congruent to positive residues modulo $3$). This is proved in Supplementary material~\cite{Note1} by using triangle inequality and positivity of trace distance. Thus, the $3$-outcome POVMs are jointly measurable when
\begin{equation}
\sum_{i,j=1}^3 | \vec{a}_i+\vec{b}_j - \vec{\theta}^s_{ij} | \leq 9.
\end{equation}
For $|\vec{a}|=|\vec{b}|=\mu$, this criterion is reduced to $\mu \le \mu_{\text{th}} := \sqrt{3}/[\sin(\phi/2)+\sqrt{3}\cos(\phi/2)]$. In the case, the unsharpness of POVM $A$ equals that of $B$, ${\cal R}_A = {\cal R}_B = {\cal R}$. Let ${\cal R}_\text{th}$ be a threshold unsharpness that, for ${\cal R} \ge {\cal R}_\text{th}$, the POVMs $A$ and $B$ are jointly measurable. The threshold unsharpnesses ${\cal R}_\text{th}$ and $\mu_\text{th}$ are presented in Table.~\ref{table:3o2d} for some relative angles. The results show that ${\cal R}_\text{th}$ increases, as one triangle rotates away from the other, in other words, as one POVM is more incompatible to the other (see Fig.~\ref{fig:RPthree}).

{\em Selective and Sequential Measurements.---}JM and non-JM can operationally be tested by using the operational quasiprobability (OQ)~\cite{Ryu2013,Jae2017}. OQ is constructed with probabilities by the measurements, $A$, $B$, and $C$: $Q(i,j) = p_C(i,j)+(p_A(i)-\sum_j p_C(i,j))/2+(p_B(j)-\sum_i p_C(i,j))/2$, where $p_X(x)$ is a probability of obtaining outcome $x$ by measurement $X$. In quantum theory, the OQ is the quantum expectation of the $W$-measure for a quantum state $\hat{\varrho}$, $Q(i,j) = \text{Tr} \hat{W}_{ij} \hat{\varrho}$.

In particular we adopt a sequential measurement as a conjunction instead of exploring all possible conjunctions. This leads that negative value of $Q(i,j)$ for some $\hat{\varrho}$ is a necessary condition for non-JM (reciprocally, sufficient for JM) and enables to stay away from a minimization in Eq.~\eqref{eq:negativity}. The conjunction POVM $C$ is sequential of $A$ and $B$ that a system is measured by POVM $A$ first, whose output state is measured by POVM $B$ later~\cite{Note2}. The conjunction POVM $C$ is given in a form of 
\begin{eqnarray}
\label{eq:seqm}
\hat{C}_{ij}=\hat{K}_{i}^\dag \hat{B}_j \hat{K}_{i},
\end{eqnarray}
where $\hat{K}_i$ are Kraus operators for $A$ with $\hat{A}_i = \hat{K}_i^\dag \hat{K}_i$. Note that the conjunction probability $p_C(i,j) = \text{Tr} \hat{C}_{ij} \hat{\varrho}=\text{Tr} \hat{B}_j \left( \hat{K}_i \hat{\varrho} \hat{K}_i^\dag \right) $. The non-JM test now involves the sequential measurement, $C$, and the individual $A$ and $B$. We call these {\em selective and sequential measurements} (SSM)~\cite{Ryu2013,Jae2017}.

Our test in the SSM scenario is also sufficient for non-JM in case of the $2$-outcome unbiased measurements $A$ and $B$. Assuming $\hat{K}_{i} = \sqrt{\hat{A}_i}$, the $W$-measure is given by
\begin{equation}
\hat{W}^S_{ij}=\frac{1}{4}\left[ \left(1+\omega^{i+j}\vec{a}\cdot\vec{b} \right){\mathbb{1}} + \left(\omega^i\vec{a}+\omega^j\vec{b}\right)\cdot\vec{\sigma}\right].
\end{equation}
It is positive when $1 \pm \vec{a}\cdot\vec{b} - | \vec{a} \pm \vec{b} |\geq0$, $\forall i,j$, and this is equivalent to the criterion~\cite{Busch86}, $| \vec{a} + \vec{b} | + | \vec{a} - \vec{b} | \leq 2$. Thus, the positivity condition of the OQ (or that of $W$-measure) fully and operationally characterizes the JM of $2$-outcome unbiased POVMs in the SSM scenario.

The test of OQ also identifies violation of macrorealism~\cite{Leggett85,Clemente15} and measurement-selection context~\cite{Spekkens05} in given systems~\cite{Ryu2013,Jae2017,Ryu2018}. OQ can pave the way to reveal relations among non-JM and the quantum features, which is beyond this work.

{\em Conclusions.---}We provided the systematic method to operationally examine the joint measurability by introducing the $W$-measure. We showed that the positivity of $W$-measure is the necessary and sufficient condition for the joint measurability between two polychotomous variables on a qu$D$it. To illustrate our method, we derived the new criterion of joint measurability for the trichotomous variables. We also suggested the operational test for the non-joint measurability by the selective and sequential measurement. Our method can be applied to detection of steerability~\cite{Uola2014,*Quintino2014}. It is a forthcoming work to generalize the present work to more-than-two variables~\cite{Ryu2013} and/or continuous variables~\cite{Jae2017}.

\begin{acknowledgments}
We thank Wonmin Son for discussions. This research was supported by the National Research Foundation of Korea (NRF) grants (No.2014R1A2A1A10050117 and No.2019R1A2C2005504), funded by the MSIP (Ministry of Science, ICT and Future Planning), Korea government, and supported by the MSIT(Ministry of Science and ICT), Korea, under the ITRC(Information Technology Research Center) support program(IITP-2019-2015-0-00385) supervised by the IITP(Institute for Information \& communications Technology Promotion). JR acknowledges the National Research Foundation, the Prime Minister's Office, Singapore and the Ministry of Education, Singapore under the Research Centres of Excellence programme.
\end{acknowledgments}


\begin{appendix}
\renewcommand{\thesection}{S\arabic{section}}
\renewcommand{\theequation}{S\arabic{equation}}
\bigskip  
\centerline{\Large{Supplementary Material}}
{\section{S1. proofs of theorems 1 and 2, and corollary 1}}

{\em Proof of Theorem 1.---}We first prove `only if' part. If the $W$-measure is positive for the measurements $(\hat{A}_i, \hat{B}_j, \hat{C}_{ij})$, it is a joint POVM for $A$ and $B$ since $\sum_j \hat{W}_{ij} = \hat{A}_i$ and $\sum_i \hat{W}_{ij} = \hat{B}_j$ hold for all $i$ and $j$ by definition. The `if' part is proved by employing a joint POVM $J$ as the conjunction $C_{ij}$. If $A$ and $B$ are jointly measurable and their joint POVM is $J=\{ \hat{J}_{ij}\}$, then the POVMs $(\hat{A}_i, \hat{B}_j, \hat{J}_{ij})$ always result the positive $W$-measure. By the marginality of the joint POVM $\sum_j \hat{J}_{ij} = \hat{A}_i$ and $\sum_i \hat{J}_{ij} = \hat{B}_j$, the $W$-measure becomes POVM ${J}$.\hspace*{0pt}\hfill\qedsymbol

{\em Proof of Corollary 1.---} Suppose that $W$-measure is obtained by the two POVMs, $A$, $B$, and conjunction operator $\{\hat{C}'_{ij}\}$. The $\hat{C}'_{ij}$ is assumed as an arbitrary operator allowed to be negative and satisfying $\sum_{ij}\hat{C}'_{ij}={\mathbb{1}}$. A set of the arbitrary operators is not POVM. Let such $W$-measure is written by $\hat{W}_{ij}(\hat{C}'_{ij})$. If such $W$-measure is a POVM, one can always find another POVMs, $(\hat{A}_i,\hat{B}_j, \hat{W}_{ij}(C'_{ij}) )$, which result the same positive $W$-measure. The $W$-measure obtained from these three POVMs result $\hat{W}_{ij}(\hat{W}_{ij}(\hat{C}'_{ij})) = \hat{W}_{ij}(\hat{C}'_{ij})$ by the marginality of $\hat{W}_{ij}(\hat{C}'_{ij})$, $\sum_i\hat{W}_{ij}(C'_{ij}) = \hat{B}_j$ and $\sum_j\hat{W}_{ij}(C'_{ij}) = \hat{A}_i$. Thus, a positive $W$-measure is always obtained by three POVMs including two POVMs $A$, $B$, and conjunction POVM $C$.\hspace*{0pt}\hfill\qedsymbol

{\em Proof of Theorem 2.---} If $\hat{W}(\Theta)$ is positive and ${\cal N}(\Theta)=0$ for some $\Theta$, the test POVM $A$ and $B$ are jointly measurable. If $A$ and $B$ are jointly measurable, their joint POVM is always written in the form of Eq.~(1) as shown in Corollary 1, and $\hat{\Theta}_{ij}$ is given by $\frac{1}{d}\sum_j \hat{C}_{ij}+\frac{1}{d}\sum_i \hat{C}_{ij}-\hat{C}_{ij}$. Thus, there exists $\Theta$ which results ${\cal N}(\Theta)=0$.\hspace*{0pt}\hfill\qedsymbol

\section{S2. Negativity of unbiased measurements}

For $D$-dimensional system, the negativity $\cal N$ is written
\begin{eqnarray}
{\cal N}&=& \frac{1}{D}\sum_{i,j} \left\Vert |\hat{W}_{ij}| - \hat{W}_{ij} \right\Vert \nonumber \\
&=& \frac{1}{D}\sum_{i,j} \Vert \hat{W}_{ij} \Vert -1 \geq0.
\end{eqnarray}
The trace norm satisfies triangle inequality, homogeneity, and convexity~\cite{Nielsen2011}.

For the $d$-outcome unbiased measurements $\hat{A}_i$ and $\hat{B}_j$, $W$-measure is given by
\begin{eqnarray}
\hat{W}_{ij}=\hat{X}_{ij} - \hat{\Theta}_{ij},
\end{eqnarray}
where $\hat{\Theta}_{ij}$ is a differential operator with constraints $\sum_i\hat{\Theta}_{ij} =  \sum_j\hat{\Theta}_{ij} =  \mathbb{1}/d$. The eigenvalues of the $W$-measure are $\lambda_k^{ij}=(2-\theta_0^{ij}+\omega^k | \vec{a}_i +\vec{b}_j - \vec{\theta}_{ij}| )/d^2$ and $\lambda_1\leq \lambda_2$. For $2$-outcome measurements, $\vec{a}_i = \omega^i \vec{a}$ and $\vec{b}_j = \omega^j \vec{b}$.

\subsection{A. 2-outcome measurements}

For $W$-measure of $2$-outcome measurements, the negativity becomes
\begin{eqnarray}\label{eq:2N}
{\cal{N}} = \frac{1}{2} \sum_{i,j=1}^2 \Vert \hat{X}_{ij} - \hat{\Theta}_{ij} \Vert-1,
\end{eqnarray}
where $\hat{X}_{ij}= (\hat{A}_i + \hat{B}_j)/2$. The $\Theta$ constraints, $\sum_i \hat{\Theta}_{ij} = \sum_j \hat{\Theta}_{ij}=\openone/2$, allow us to choose $\hat{\Theta}_{11}$ for an arbitrary free (Hermitian) operator. The negativity is lower bounded,
\begin{eqnarray}\label{eq:2tri}
{\cal{N}}&\geq& \frac{1}{2} \left( \Vert \hat{X}_{11}-\hat{X}_{22} \Vert +  \Vert \hat{X}_{12}-\hat{X}_{21}  \Vert \right)-1,\forall\hat{\Theta}_{11}. 
\end{eqnarray}
Here we applied the triangle inequality, $\Vert \hat{Y} \Vert + \Vert\hat{Z} \Vert \geq \Vert \hat{Y}-\hat{Z} \Vert$, consecutively. The equality holds and the lower bound is attainable when $\vec{\theta}_{ij}=\vec{0}$, $\hat{\Theta}_{ij} = {\theta_0^{ij}}\mathbb{1}/4$, $\lambda_1^{ij}\le 0$, and $\lambda_2^{ij}\geq0$, $\forall i,j$. At this condition, the minimum value of negativity is given by the lower bound and its value can be explicitly reads
\begin{eqnarray}
\min_{\{ \hat{\Theta}_{ij}\}}{\cal{N}} &=&  \frac{1}{2} \left( \Vert \hat{X}_{11}-\hat{X}_{22} \Vert +  \Vert \hat{X}_{12}-\hat{X}_{21}  \Vert \right)-1 \nonumber\\
&=&\frac{1}{2} \left( |\vec{a}+\vec{b}| + |\vec{a}-\vec{b}| \right)-1.
\end{eqnarray}
Note that the condition for $\theta_0^{ij}$ is imposed on the conditions of the eigenvalues, and it is reduced to $2-|\vec{a} + \vec{b} | \leq \theta_0^{11} \leq |\vec{a}-\vec{b}|$ by the $\hat{\Theta}$ contraints. Such $\theta_0^{11}$ can exist when $2-|\vec{a} + \vec{b} | \leq |\vec{a} - \vec{b} |$. If the equality in Eq.~\eqref{eq:2tri} does not hold, $2-|\vec{a} + \vec{b} | > |\vec{a} - \vec{b} |$. This holds if and only if there exists $\theta_0^{11}$ such that $2-|\vec{a} + \vec{b} | > \theta_0^{11} > |\vec{a} - \vec{b} |$, for which $\lambda_1^{ij}>0$ and $\lambda_2^{ij}>0$, $\forall i,j$. In other words, if the equality does not hold, the $\hat{W}_{ij}>0$, $\forall i,j$, and ${\cal N}=0$.

Finally, the minimum value of negativity becomes
\begin{equation}
{\cal N}_\text{min} = \max \left[\frac{1}{2} \left( |\vec{a}+\vec{b}| + |\vec{a}-\vec{b}| \right)-1,0 \right].
\end{equation}
The negativity is zero when $ |\vec{a}+\vec{b}| + |\vec{a}-\vec{b}| \leq 2$, and this condition is equivalent to Busch's joint measurability criterion~\cite{Busch86}. \hspace*{0pt}\hfill\qedsymbol

\subsection{B. 3-outcome measurements}

For $3$-outcome measurements, the negativity becomes
\begin{equation}
{\cal N} = \frac{1}{2}\sum_{i,j=1}^3 \Vert \hat{X}_{ij} - \hat{\Theta}_{ij} \Vert -1.
\end{equation}
The four differential operators are free parameters to consider their joint measurability problem of $3$-outcome measurements. We select $\hat{\Theta}_{11},~ \hat{\Theta}_{22},~ \hat{\Theta}_{+}=\hat{\Theta}_{12}+\hat{\Theta}_{21}$, and $\hat{\Theta}_{-}=\hat{\Theta}_{12}-\hat{\Theta}_{21}$ as the free parameters.

\begin{figure}[t]
	\centering
	\includegraphics[width=0.35\textwidth]{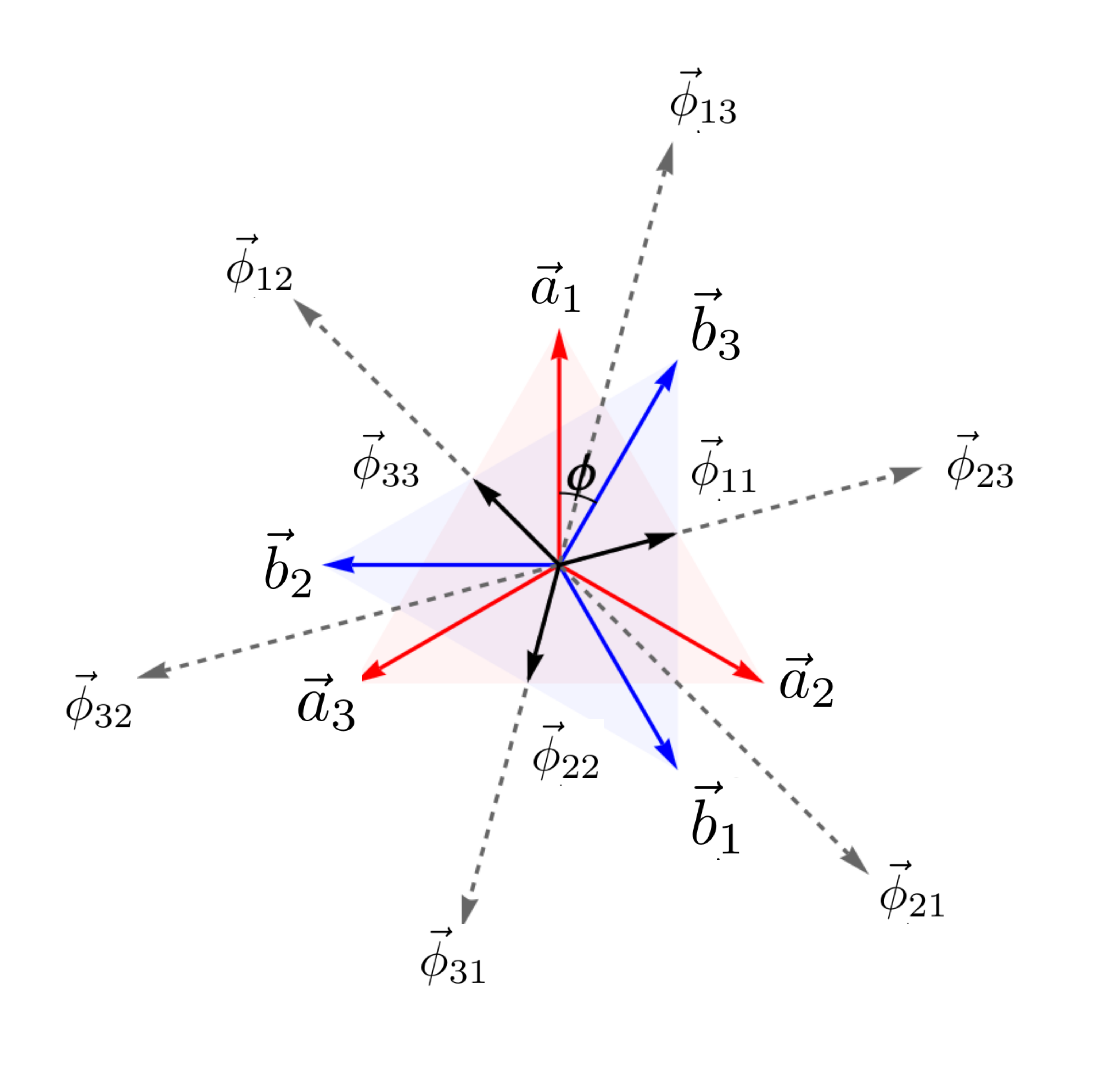}
	\caption{Geometry of Bloch vectors of trichotomous POVM $A$ and $B$. Relative angle between Bloch vectors $\{ \vec{a}_i\}$ and $\{ \vec{b}_j\}$ is $\phi$. $\vec{\phi}_{ij}$ is defined by $\vec{\phi}_{ij} = \vec{a}_i + \vec{b}_j$. The vectors $\vec{\phi}_{ii}$, $\vec{\phi}_{i+1i+2}$ and $\vec{\phi}_{i+2i+1}$ are parallel.}
\label{fig:geo}
\end{figure}
By using the triangle inequality and the positivity of trace norm, we can obtain the lower bound of the negativity function:
\begin{eqnarray}\label{eq:3tri}
{\cal N}&\geq & \frac{1}{2} \left( \Vert \hat{X}_{11} - \hat{\Theta}_{11} \Vert + \Vert \hat{X}_{22} - \hat{\Theta}_{22} \Vert + \Vert \hat{X}_{33}-\hat{\Theta}_{33}\Vert \right) \nonumber\\
&+&\frac{1}{2} \sum_{i=1}^3 \Vert \hat{X}_{ii+2} - \hat{X}_{i+2i} - \hat{\Theta}_{-}\Vert -1   \nonumber\\
&\geq& \frac{1}{2} \sum_{i=1}^3 \Vert \hat{X}_{ii+2} - \hat{X}_{i+2i} - \hat{\Theta}_{-}\Vert -1 \nonumber\\
&\geq& \frac{1}{2} \sum_{i=1}^3 \Vert \hat{X}_{ii+2} - \hat{X}_{i+2i} \Vert -1, ~\forall \hat{\Theta}_{11}, \hat{\Theta}_{22}, \hat{\Theta}_{+}, \hat{\Theta}_{-}, \nonumber\\
\end{eqnarray}
where $\hat{\Theta}_{33}=\hat{\Theta}_{11} + \hat{\Theta}_{22} + \hat{\Theta}_{+} - {\mathbb{1}}/{3}$. The minimum of the lower bound is attainable and the third equality in Eq.~\eqref{eq:3tri} holds when $\hat{\Theta}_{-}=\hat{0}$. It is determined by the solution of the problem finding operator $\hat{\Theta}_{-}$ which results minimum sum of distances from three operators $\hat{X}_{13}-\hat{X}_{31}$, $\hat{X}_{21}-\hat{X}_{12}$, and $\hat{X}_{32}-\hat{X}_{23}$. This problem is a Fermat-Toricelli (FT) point problem~\cite{OH2013} on the operator. The FT problem is finding a point which minimizes sum of distances between vertices and the point. In our problem, the difference of Bloch vectors, $\vec{\phi}_{ii+2}-\vec{\phi}_{i+2i}$, represent the operators $\hat{X}_{ii+2}-\hat{X}_{i+2i}$, and they form a regular triangle of which center point is $\vec{0}$ (see Fig.~\ref{fig:geo}). For vertices of the regular triangle, one has center point as the FT point. Similarly, the lower bound of negativity is minimized when $\hat{\Theta}_{-}=\hat{0}$, and this result reflects the FT point. The minimum value of the negativity function can be put as
\begin{equation}
\min_{\{ \hat{\Theta}_{ij} \} }{\cal N} = \frac{1}{2} \sum_{i=1}^3 \Vert \hat{X}_{ii+2} - \hat{X}_{i+2i}\Vert -1.
\end{equation}
The second equality in Eq.~\eqref{eq:3tri} holds when $\hat{\Theta}_{ii}=\hat{X}_{ii}$ or, equivalently, $\theta_0^{ii}=2$ and $\vec{\theta}_{ii}=\vec{\phi}_{ii}$, $\forall i$. This implies $\lambda_{k}^{i=j}=0$, $\forall i,k$.

The minimum value is attainable and the first equality in Eq.~\eqref{eq:3tri} holds by the solution differential operators $\hat{\Theta}_{ij}^s = (\theta_0^{ij}\mathbb{1} + \vec{\theta}_{ij}^s\cdot \vec{\sigma})/9$, where the $\vec{\theta}^s_{ij} = \vec{a}_{2(i+j)} + \vec{b}_{2(i+j)}$, i.e.,
\begin{equation}\label{M:solution}
\begin{pmatrix} 
\vec{\theta}_{11}^s & \vec{\theta}_{12}^s & \vec{\theta}_{13}^s\\
\vec{\theta}_{21}^s & \vec{\theta}_{22}^s & \vec{\theta}_{23}^s\\
\vec{\theta}_{31}^s & \vec{\theta}_{32}^s & \vec{\theta}_{33}^s\\
\end{pmatrix}
= 
\begin{pmatrix} 
\vec{\phi}_{11} & \vec{\phi}_{33} &  \vec{\phi}_{22}\\
\vec{\phi}_{33} & \vec{\phi}_{22} & \vec{\phi}_{11}\\
\vec{\phi}_{22} & \vec{\phi}_{11} & \vec{\phi}_{33}\\
\end{pmatrix}
,
\end{equation}
with $\lambda_k^{i=j}=0$, $\lambda_1^{i\neq j}\leq 0$ and $\lambda_2^{i\neq j}\geq0$, $\forall i,j,k$. The conditions for $\theta_0^{ij}$ are imposed on the conditions of eigenvalues. They are reduced to $\theta_0^{11}=\theta_0^{22}=2$, $\theta_0^{+} = 1$, and
\begin{eqnarray}
4-|\vec{\phi}_{12} - \vec{\phi}_{33}| -|\vec{\phi}_{21} - \vec{\phi}_{33}| \leq &\theta^+_0& \leq 4 + |\vec{\phi}_{12} - \vec{\phi}_{33}| + |\vec{\phi}_{21} - \vec{\phi}_{33}|, \nonumber\\
-|\vec{\phi}_{31} - \vec{\phi}_{22}| -|\vec{\phi}_{13} - \vec{\phi}_{22}| -2 \leq &\theta^+_0& \leq  |\vec{\phi}_{31} - \vec{\phi}_{22}| + |\vec{\phi}_{13} - \vec{\phi}_{22}| -2, \nonumber\\
-|\vec{\phi}_{23} - \vec{\phi}_{11}| -|\vec{\phi}_{32} - \vec{\phi}_{11}| -2 \leq &\theta^+_0& \leq  |\vec{\phi}_{23} - \vec{\phi}_{11}| + |\vec{\phi}_{32} - \vec{\phi}_{11}| -2, \nonumber
\end{eqnarray}
where $\vec{\phi}_{ij} = \vec{a}_i + \vec{b}_j$. [Note that the conditions, $\theta_0^{11}=\theta_0^{22}=2$, $\theta_0^{+} = 1$, determine the second equality in Eq.~\eqref{eq:3tri}, and the conditions of the above three inequalities are concerned with the first inequality in Eq.~\eqref{eq:3tri}.] The $\theta_0^{+}=1$ if $|\vec{\phi}_{ii+1}-\vec{\theta}^s_{ii+1}| + |\vec{\phi}_{i+1i}-\vec{\theta}^s_{i+1i}| \geq 3$, $\forall i$. If the first equality does not hold, $|\vec{\phi}_{ii+1}-\vec{\theta}^s_{ii+1}| + |\vec{\phi}_{i+1i}-\vec{\theta}^s_{i+1i}| < 3$ for some $i$. The latter holds when $\theta_0^+=1$ such that 
\begin{eqnarray}
&&\theta^+_0 < 4-|\vec{\phi}_{12} - \vec{\phi}_{33}| -|\vec{\phi}_{21} - \vec{\phi}_{33}|, \nonumber\\
\text{or}~&&\theta^+_0 >  |\vec{\phi}_{31} - \vec{\phi}_{22}| + |\vec{\phi}_{13} - \vec{\phi}_{22}| -2,\nonumber\\
\text{or}~&&\theta^+_0 >  |\vec{\phi}_{23} - \vec{\phi}_{11}| + |\vec{\phi}_{32} - \vec{\phi}_{11}| -2.\nonumber
\end{eqnarray}
These are equivalent to $\lambda^{i=j}_k =0$, $\lambda^{i\neq j}_1 > 0$, and $\lambda^{i\neq j}_2 >0$, for which $\hat{W}_{ij} \geq 0$ $\forall i,j,k$, and ${\cal N}=0$.

Finally, the negativity $\cal N$ with $\hat{\Theta}_{ij}^s$ can be written by
\begin{equation}\label{eq:threN}
{\cal N}_\text{min} = \max \left( \frac{1}{9}\sum_{i,j=1}^3 |\vec{a}_i+\vec{b}_j - \vec{\theta}^s_{ij}| - 1, 0 \right).~~ \text{\hspace*{0pt}\hfill\qedsymbol}
\end{equation}
\end{appendix}

\end{document}